\documentclass[12pt]{article}
\renewcommand\({\left(}
\renewcommand\){\right)}

\newcommand\eq[1]{Eq.~(\ref{#1})}
\newcommand\eqs[2]{Eqs.~(\ref{#1}) and (\ref{#2})}

\newcommand\ee{\end{equation}}
\newcommand\be{\begin{equation}}
\newcommand\eea{\end{eqnarray}}
\newcommand\bea{\begin{eqnarray}}

\newcommand\mpl{M_{\rm P}}

\newcommand\lsim{\mathrel{\rlap{\lower4pt\hbox{\hskip1pt$\sim$}}
    \raise1pt\hbox{$<$}}}
\newcommand\gsim{\mathrel{\rlap{\lower4pt\hbox{\hskip1pt$\sim$}}
    \raise1pt\hbox{$>$}}}

\newcommand\diff{\mbox d}

\newcommand\calp{{\cal P}}
\newcommand\calr{{\cal R}}
\newcommand\calpr{{\calp_\calr}}

\newcommand\sub[1]{_{\rm #1}}

\begin{document}
\date{}
\title{The primordial curvature perturbation\\ in the ekpyrotic Universe}

\author{David H.~Lyth
\\
\\
Physics Department, Lancaster University, Lancaster LA1 4YB,  U.K.}
\sloppy
\maketitle

\begin{abstract}\noindent
In the ekpyrotic scenario the Universe is initially collapsing,  the energy 
density coming from a scalar field with a negative exponential potential. 
On the basis of a calculation ignoring the gravitational back-reaction the 
authors of the scenario claim that during collapse the vacuum fluctuation 
creates a perturbation in the comoving curvature, which has  a flat spectrum 
in accordance with observation. In this note the back-reaction is included, 
and it is found that the spectrum during collapse is strongly scale-dependent 
with negligible magnitude. The spectrum is continuous across the bounce
if the spacetime is smooth, making it unlikely that the ekpyrotic scenario
can be compatible with observation.

\end{abstract}

 \paragraph{Introduction}
In the ekpyrotic Universe  \cite{ek,ek2} (called in a slightly
different version \cite{py} the pyrotechnic Universe)
the Big Bang originates when
a brane moving in an extra dimension collides with a fixed one.
Except around the time of the collision, the extra dimension can be
integrated out to give a four-dimensional field  theory with Einstein gravity
and flat space. Before the
collision, the Universe is collapsing, with the 
energy density provided by the  field $\phi$ which defines
the position of the moving brane. In this phase the
Hubble parameter $H$, the energy density $\rho$ and the negative potential
$V$ are related by the Friedmann equation
\be
\rho=3\mpl^2H^2 =\frac12 \dot\phi^2 + V(\phi)\label{1}
\,.
\ee
The Universe is supposed initially to be almost static with
negligible energy density. The  energy density
increases as the Universe collapses, while remaining negligible compared with
the potential.
When the moving brane collides with the stationary one, the energy
density is converted to radiation, and at the same time
the Universe starts to expand.

This ekpyrotic scenario is similar to the pre-big-bang scenario 
\cite{pbb,davidref,clw}. 
In both cases the 4-D field theory breaks down at the bounce,
whose description would in principle require a string theory calculation
(though the calculation may be more tractable in the ekpyrotic case since
the bounce can take place far below the string scale).
The main difference is that in the pre-big-bang scenario 
 the scalar fields responsible for the energy density during collapse 
are supposed to have negligible potential.

The adiabatic density perturbation responsible for structure
in the Universe is conveniently characterized by the curvature
perturbation $\calr$ seen by comoving observers 
\cite{bardeen,areview,book,treview}. 
Its  spectrum
$\calpr$ 
must be practically scale-independent  (flat).
A flat  spectrum
is generated during inflation with a sufficiently flat potential,
and existing calculations \cite{ek,py} seem to indicate
that it is also generated
 during ekpyrotic collapse
 provided that the potential has a string-inspired
exponential form,
\be
V=-V_0\exp\(-\sqrt\frac2 p \frac\phi\mpl\) \label{5}
\,.
\ee
(In \cite{ek,py}, the potential was actually taken to be an exponential
function of a field with slightly non-canonical normalization, leading to
a slightly scale-dependent spectrum, but for simplicity 
we focus on the case where
the canonically-normalized field appears in the potential.)
However, these calculations ignore gravitational back-reaction.
In this  note we include the back-reaction and obtain 
 a quite different result. 

 \paragraph{The unperturbed Universe}
The case where the Universe is dominated by a field with
an exponential potential has already been studied in the context of
power-law inflation \cite{book,ls}. We focus first on the unperturbed
case. In addition to the Friedman equation  one needs the field equation
\be
\ddot\phi+3H\dot\phi + V'=0 \label{6}\,,
\ee
or equivalently the relation
\be
\mpl^2\dot H =-\frac12\dot\phi^2 \label{9}
\,.
\ee
With an exponential potential there is an exact solution of \eqs{1}{9},
corresponding to 
\bea
a&\propto& |t|^p\\
H&=&p/t\\
\frac12\dot\phi^2 &=& \frac1 p\mpl^2H^2\\
V&=&\frac{3p-1}p\mpl^2H^2 
\,.
\eea
We will also need 
the conformal time defined by $\diff\tau=\diff t/a$, which can be taken as
\be
\tau=  \frac p{p-1} \frac1{aH} 
\,.
\ee

Choosing $t>0$ and $p>1$ gives power-law inflation.
Choosing $t<0$ and $p=\frac13$ give the pre-big-bang scenario in which
there is collapse with  no potential. Choosing  $t<0$
and $p\ll \frac13$ gives  the ekpyrotic scenario in which there is collapse
with a dominant potential. As
 it is hardly more effort we deal with the general case
$0<p<\frac13$ giving collapse with a potential of arbitrary magnitude.
In the ekpyrotic case,
\bea
\rho&\ll& \dot\phi^2\\
\frac12\dot\phi^2+ V&\simeq& 0 \label{2}\\
\ddot\phi + V' &\simeq & 0 \label{2a}\\
a &\simeq& {\rm const}
\,.
\eea

 \paragraph{The curvature perturbation generated during inflation}
Let us first summarize the way in which the vacuum fluctuation 
of the inflaton field $\phi$  generates the curvature
perturbation
 \cite{areview,book,treview}.
Each perturbation $\delta\phi$ with  wavenumber $k/a$ evolves independently,
and using conformal time the quantity $u\equiv a \delta\phi$
 has the same dynamics as a free field living
in flat spacetime, with wavenumber $k$ but with some time-dependent 
mass-squared. Well before horizon exit at $aH=k$
 the mass is negligible,  and 
 at the classical level the perturbation is then supposed to 
vanish (no particles). There is however a vacuum fluctuation,
which after horizon exit   becomes
a classical gaussian  perturbation $\delta\phi$. Evaluated 
on spatially flat slices, it determines the curvature perturbation
$\calr$  seen by comoving observers \cite{sasaki},
\be
\calr= - H\delta\phi/\dot\phi \label{15}
\,.
\ee
This quantity is constant during  inflation,  and 
  on the usual
assumption that the pressure perturbation is adiabatic it
remains  constant until  horizon entry. For any sufficiently flat
potential the spectrum of $\calr$ is flat in accordance with observation.

 \paragraph{The calculation ignoring the metric perturbation}

Next we summarize the calculation presented in \cite{ek,py},
working therefore with the case $p\simeq 0$ 
The
essential assumption of that  calculation
is that the metric perturbation may be ignored, leading to\footnote
{The authors of \cite{ek} use the 
Guth-Pi-Olson (GPO) formalism \cite{olson,gp} for  cosmological perturbations,
which however is easily translated to the more standard formalism
used here. Indeed, the GPO variable $S$ is \cite{olson} $\frac23(k/aH)^2
\calr$  and this quantity immediately after
the bounce  \cite{ek} (or inflation \cite{gp}) is equated with
$-\frac23 (k/aH)^2 H\delta\phi/\dot\phi$ evaluated just before the bounce.
Comparing with \eq{15} we see that the $\delta\phi$ of the GPO formalism
is evaluated on spatially flat slices. We see also that the use of the
GPO formalism requires  $\calr$ to be
 continuous across the bounce, which in contrast with the case of inflation
is a non-trivial assumption.}
\be
\ddot{\delta\phi} + [(k/a)^2 + V''] \delta\phi=0  \label{17}
\,.
\ee
This can be written 
\be
\ddot{\delta\phi}  + \(\frac{k^2}{a^2}-\frac2{t^2}\) 
\delta\phi=0  \label{20}
\,.
\ee

We  give first a rough  argument \cite{py}. The amplification of the 
quantum fluctuation  takes place around the epoch
$(k/a)^2=|V''|$. Indeed, well before this epoch 
 the quantum fluctuation is that of a massless free 
field in flat spacetime, which on dimensional grounds 
has the spectrum $\calp_\phi\sim (k/a)^2$. For an estimate, we take this 
expression to be
valid also at the epoch $(k/a)^2=|V''|$. 
Well after this epoch,  the negative mass-squared
$V''$ dominates, and $\delta\phi$ increases.
In fact, $\delta\phi\propto 1/t\propto \dot\phi$,
 so that 
\be
\calp_\phi/\dot\phi^2 \sim |V''/V| = 2/ p\mpl^2  \label{19}
\,.
\ee
This gives $\calpr\sim (2/p)(H/\mpl^2)^2$, which when evaluated
at reheating is supposed \cite{ek,py}
to give the primordial curvature perturbation
for the subsequent hot big bang. 

To obtain a more precise result, we 
go  to 
 the quantity $u\equiv a\delta\phi$, and conformal time
$\tau$. Since we are working in the limit $p\simeq 0$
where $a$ can be taken to be constant, this change is trivial,
\eq{9} becoming
\be
\frac{\diff^2 u}{\diff \tau^2}  + (k^2-2/\tau^2) u=0  \label{20a}
\,.
\ee
This is the same equation as for slow-roll inflation, and using flat
spacetime field theory to define the initial condition at
$\tau^2\gg k^2$ one finds   \cite{sasaki,book,treview}
at $\tau^2 \ll k^2$ a  perturbation with
spectrum
\be
\calp_u = (2\pi \tau)^{-2}  \label{21}
\,.
\ee
In the case of slow-roll inflation, $\tau=-1/(aH)$ leading to
$\calp_\phi = (H/2\pi)^2$
and 
\be
\calpr= (H/\dot\phi)^2(H/2\pi)^2  \label{23}
\,.
\ee
In our case, $\tau\simeq t/a$  leading to
$\calp_\phi= (2\pi t)^{-2}$
and 
\be
\calpr =\frac1{8\pi^2 p} \(\frac {H}{\mpl}\)^2  
\label{25}
\,.
\ee

\paragraph{The calculation including the metric perturbation}

In the case of slow-roll inflation the 
metric perturbation is indeed negligible compared with the field perturbation.
In general though, they are of the same order. Using the spatially flat
slicing,  $u\equiv a\delta\phi$ satisfies \cite{sasaki}
\be
\frac{\diff^2 u}{\diff \tau^2}  + \(k^2-\frac1z \frac{\diff^2 z}{\diff \tau^2}
\) u=0  \label{26}
\,,
\ee
where 
\be
z=a\dot\phi/H  \label{26a}
\ee

For a practically static Universe, $z\propto\dot\phi$, which using
\eq{2a} gives \eq{17}. For
slow-roll inflation, $z\propto a\propto 1/\tau$
giving \eq{20a}. In our case, $z\propto t^p\propto \tau^\frac p{1-p}$ giving
 \cite{ls},
\be
\frac{\diff^2 u}{\diff \tau^2}  + \(k^2+
\frac{p(1-2p)}{(1-p)^2}\tau^{-2}\) u =0
\label{ueq}
\,.
\ee
Power-law inflation corresponds to $p>1$, giving  a negative 
mass-squared which amplifies the quantum fluctuation \cite{ls}. The collapsing
case we are considering  corresponds to $0<p<\frac13$, giving a positive 
mass-squared which does not amplify the quantum fluctuation.

We could obtain an estimate by repeating the previous rough argument,
but choose instead to go straight to the precise result. We are interested
in the late-time era when $k^2$ is negligible, and the spectrum is then
 \cite{ls} 
\be
\calpr^\frac12 = \frac{r(p)}{2\pi} \sqrt\frac p 2
 \(\frac{H}{\mpl} \) 
\(\frac k{aH} \)^\frac1{1-p}
\,.
\ee
{}In contrast with the earlier result \eq{25},
this  correctly-calculated quantity is time-independent 
(because $a\propto t^p$). In particular, 
\be
\calpr^\frac12 =  \frac{r(p)}{2\pi} \sqrt\frac p 2
 \(\frac{H\sub{reh}}{\mpl} \) 
\(\frac k{a\sub{reh}H\sub{reh}} \)^\frac1{1-p}
\,.
\ee
On cosmological scales  $k/a\sub{reh}H\sub{reh}\sim e^{-N}$
with $N\lsim 60$, the precise value depending on $H\sub{reh}$ and
the subsequent cosmology \cite{book,treview}. 
The  spectral index is
\be
n=1+\frac2{1-p}
\,.
\ee
which gives $n=3$ for the ekpyrotic case and reproduces the known result
$n=4$ for the pre-big-bang case.

\paragraph{The bounce and beyond}

We shown that the curvature perturbation during collapse is unviable,
but it needs to be evolved to the epoch of horizon entry before one can
say that the ekpyrotic (or pre-big-bang)  scenario is unviable.
The first step it to go across the bounce. The continuity of $\calr$
is assumed in pre-big-bang discussions and we have seen that it was also
assumed in the ekpyrotic scenario \cite{ek,py}. In a separate paper 
\cite{mynew} it is demonstrated that  $\calr$ is indeed continuous in the
approximation that the bounce occurs on a unique slice of spacetime.
Also, its near-constancy is maintained just after the bounce
provided that this slice is smoothly  embedded in spacetime. 

The validity
of these approximations can be ascertained only by a calculation going beyond
the 4-D framework. Such a calculation may well  violate the assumptions
and generate a curvature perturbation, but there is no reason to suppose
that its spectrum will be  flat.
Accepting the validity of the assumptions $\calr$ will be constant
until horizon entry, in contradiction with observation,  unless there is either
a non-adiabatic pressure perturbation or anisotropic pressure 
\cite{bardeen,areview, book,treview}. 
It may be possible to generate a non-adiabatic
pressure perturbation
through the vacuum fluctuation of  field {\em not} contributing
significantly to the energy density during collapse,
but judging experience with the pre-big-bang case  \cite{davidref,clw}
it will be difficult to make the spectrum of the fluctuation flat.
It seems fair to say that, compared with the inflationary scenario,
 the ekpyrotic and pre-big-bang scenarios are disfavored by the requirement
of structure formation.

\paragraph{Conclusion}

In the ekpyrotic scenario
  the Universe is initially contracting under the influence of an exponential
potential. We have shown that
the curvature perturbation responsible for the origin of structure
is not generated while the Universe is collapsing, and have argued that
it is unlikely to be generated subsequently. 

As was pointed out in \cite{py}, a
 simple modification of the ekpyrotic scenario
 is to generalize the string-inspired potential
\eq{5} by adding a constant to it. One then has an inflation model
\cite{gia},  which can generate a viable curvature perturbation in 
the usual way.

\section*{Acknowledgments}
I thank Paul Steinhardt,  Neil Turok and David Wands
for  discussions about this subject.


\newcommand\pl[3]{Phys.\ Lett.\ {\bf #1}  (#3) #2}
\newcommand\np[3]{Nucl.\ Phys.\ {\bf #1}  (#3) #2}
\newcommand\pr[3]{Phys.\ Rep.\ {\bf #1}  (#3) #2}
\newcommand\prl[3]{Phys.\ Rev.\ Lett.\ {\bf #1}  (#3)  #2}
\newcommand\prd[3]{Phys.\ Rev.\ D{\bf #1}  (#3) #2}
\newcommand\ptp[3]{Prog.\ Theor.\ Phys.\ {\bf #1}  (#3)  #2 }
\newcommand\rpp[3]{Rep.\ on Prog.\ in Phys.\ {\bf #1} (#3) #2}
\newcommand\jhep[2]{JHEP #1 (#2)}
\newcommand\grg[3]{Gen.\ Rel.\ Grav.\ {\bf #1}  (#3) #2}
\newcommand\mnras[3]{MNRAS {\bf #1}   (#3) #2}
\newcommand\apjl[3]{Astrophys.\ J.\ Lett.\ {\bf #1}  (#3) #2}

\end{document}